# Optically pumped nanolaser based on two magnetic plasmon resonance modes

Z. H. Zhu,[1,2] H. Liu,[1,*] S. M. Wang,[1] T. Li,[1] J. X. Cao,[1] W. M. Ye,[2] X. D. Yuan,[2] S. N. Zhu,[1,*]

[1]*Department of Physics, National Laboratory of Solid State Microstructures, Nanjing University, Nanjing 210093, People's Republic of China*

[2]*College of Opto-Electronic Engineering, National University of Defense Technology, Changsha 410073, People's Republic of China*

**Abstract**

We propose and analyze theoretically a double magnetic plasmon resonance nanolaser, in which Ytterbium-erbium co-doped material is used as the gain medium. Through design of the double magnetic resonance modes, pumping light (980nm) can be resonantly absorbed and laser light (1550nm) can be resonantly generated simultaneously. We introduce a set of rate equations combined to describe the operation of the laser and predict the lasing condition. According to our calculations, the disadvantage that pumping light is difficult to be absorbed by a thin slab of gain materials can be overcome.

[*]Email address: liuhui@nju.edu.cn; zhusn@nju.edu.cn
[*]URL: http://dsl.nju.edu.cn/dslweb/images/plasmonics-MPP.htm

Reducing mode volume in cavities is very favorable for achieving strong light-matter interaction processes.[1] These strong interactions are of great benefit to applications in light emitting devices. Recently, in order to further miniaturize mode volume and physical size of structure, researchers turn to use metal instead of dielectric to bound the mode field.[2-6] For example, a metallic-coated cavity formed by encapsulating a semiconductor heterostructure in a thin gold film is used to obtain mode volume far smaller than conventional dielectric cavities.[2] In addition, due to the strong localized surface plasmon (SP) effect, electromagnetic energy is also able to be highly confined in nano-scale.[3-6]

In 1999, Pendry proposed the split-ring resonator (SRR),[7] to realize magnetic plasmon resonance in microwaves. Since then, scientists introduce many new structures to achieve magnetic resonance in high frequency region, such as fish-net,[8] and nanosandwich.[9] It is shown that such magnetic response structures offer a great promise for applications in negative refractions,[10-12] perfect len,[13] subwavelength waveguides,[14] etc. Very recently, magnetic plasmon resonance is proposed as an efficient way to produce nanolasers.[15-16] However, to the best of our knowledge, all the structures up to now only resonate with the lasing light. In these structures, most of external energy is difficult to be absorbed by these nanocavities.

In this work, we design a metallic nanosandwich cavity, which can possess two magnetic plasmon resonance modes at two prescribed resonance wavelengthes. Combined with the Ytterbium-erbium co-doped gain material, this cavity can have resonance not only at lasing wavelength but also at pumping wavelength, which lead to efficient pump-laser conversion. In this paper, for principle demonstration, we selected the Er:Yb:YCa$_4$O(BO$_3$)$_3$ (Er:Yb:YCOB) crystal [17-20] as the gain medium, which is widely used to generate light in optical communication wavelength region around 1.5 $\mu m$。

Figure 1 (a) presents the proposed laser arrays constructed by immersing a regular array of rectangular silver slab in gain medium Er:Yb:YCOB, which is supported by a silver film and SiO$_2$ glass substrate. The geometry parameters of the structure shown in Fig. 1(b) and Fig. 1(c) are defined as the following: The immersed upper silver slab is a rectangular slab with a size of $310nm \times 245nm \times 50nm$. The gap size between the immersed silver slab and bottom silver film is 50nm. The thicknesses of the silver film, gain medium layer above the silver film and SiO$_2$ glass substrate are 50nm, 250nm and 250nm, respectively. The permittivity of the silver is given by $\varepsilon(\omega) = \varepsilon_\infty - \omega_p^2/(\omega^2 + i\omega/\tau)$. The values of $\varepsilon_\infty$, $\omega_p$ and $\tau$ fitted to experimental data in the 950–1800 nm wavelength range are 1.0, $1.38 \times 10^{16}$ rad/s, and 33 fs, respectively. The refractive indexes of Er:Yb:YCOB and SiO$_2$ substrate are measured to be 1.3 and 1.5, respectively.

In this paper, we focus on the lasing condition of a single magnetic resonator. The distance between nearest-neighbor resonators is enough large (~1000nm) and the coupling interaction among these resonators can be neglected. From Pendry's theory, the structure shown in Fig. 1(b) can be seen as an equivalent inductance-capacitance (LC) circuit. When magnetic resonance occurs, the electromagnetic energy is mainly stored in the space between the silver slab and film, so the structure can be considered as a cavity of confining light. One main character of magnetic resonance mode is that the induced current forms loops and strong magnetic fields inside them. For different magnetic resonance mode, different number of loops is formed in the cavity. So, we can label one mode as MP$_{ij}$, where i and j denote the formed loop numbers along x and y direction, respectively. It is found that the electric fields of mode $MP_{10}$ and $MP_{22}$ have quite large spatial overlap, so these two modes are selected as the operation modes in our calculations. For the carefully designed geometry size in Fig.1 (b), the resonance wavelengths of the two modes are 980nm and 1550nm respectively. Fig. 2 shows the magnetic field vector distributions of the two modes in the middle dielectric layer. Fig.3 (b) and Fig. 3(d) show respectively the electric field component Ey distributions for the two modes. As the 980nm and 1550nm are respectively close to the absorption peak and emission peak of Er:Yb:YCOB, we can select the higher order $MP_{22}$ mode as the pumping mode (980nm), and the lower $MP_{10}$ mode as the lasing mode (1550nm).

Figure 3 illustrates schematically the operation principle of the double magnetic resonance laser. To predict the lasing condition, we introduce a set of rate equations to model the operation of the laser. As we have know, if the radiative transition frequency of the atom or ion matches that of the resonance mode of a cavity, the rate of spontaneous or stimulated emission can be enhanced. The measure of the enhancement is the Purcell factor defined by $F = 3Q\lambda^3/(4\pi^2 V n^3)$,[21-22] where n, λ, Q and $V$ are the refractive index of gain material, the wavelength, quality factor, and effective mode volume of laser mode, respectively. Here, the rate equations take into account enhanced spontaneous and stimulated emission. Namely, the absorption cross section of $Yb^{3+}$ and emission cross section of $Er^{3+}$ in the rate equations are $F_l\sigma_E$ and $F_p\sigma_Y$, respectively, where $\sigma_Y$ and $\sigma_E$ are the absorption cross section of $Yb^{3+}$ and emission cross section of $Er^{3+}$ in free space (here, "free space" as the meaning of "without cavity" ). In addition, considering the localized property of the pumping and lasing modes, we also take into account the spatial distributions of two modes. The energy level diagram for the Er:Yb:YCOB is shown in Fig. 3(c). In steady-state conditions, neglecting both the populations in the levels $^4I_{11/2}$, $^4I_{9/2}$ and $^4F_{9/2}$ and corresponding back-transfer processes due to the fast nonradiative decay in these levels, the simplified rate equations and condition for threshold are expressed as [17,19]:

$$\frac{\partial N_{2Y}}{\partial t} = 0 = \sigma_Y v_p F_p N_p f_p (N_{1Y} - N_{2Y}) - k_1 N_{2Y} N_{1E} - k_2 N_{2Y} N_{2E} - \frac{N_{2Y}}{\tau_{2Y}}, \qquad (1)$$

$$\frac{\partial N_{2E}}{\partial t} = 0 = k_1 N_{2Y} N_{1E} - (N_{2E} - N_{1E}) v_l F_l \sigma_E N_l f_l - \frac{N_{2E}}{\tau_{2E}} - 2 C N_{2E}^2, \qquad (2)$$

$$\frac{\partial N_l}{\partial t} = 0 = v_l F_l \sigma_E \iiint (N_{2E} - N_{1E}) N_l f_l dV - \frac{\omega_l N_l}{Q_l}, \qquad (3)$$

where $N_{ix}$ and $\tau_{ix}$ represent respectively the population density and lifetime of the corresponding level shown in Fig. 2(c). $k_1$ and $k_2$ are the coefficients of the two energy transfer processes. C is the up-conversion rate. $\omega_l$ is the angle frequency of lasing mode. $v_p$, $N_p$ and $f_p$ are the group velocity, total photon number and normalized spatial intensity distributions of the higher order magnetic resonant mode (980nm), respectively. $v_l$, $N_l$ and $f_l$ represent respectively the corresponding parameters of the lasing mode. $f_p$ and $f_l$ are normalized such that:

$$\iiint f_p dV = 1 \text{ and } \iiint f_l dV = 1. \qquad (4)$$

Where $V$ is the volume.

In addition, in steady-state conditions, it is reasonable to give these approximately expressions:

$$N_{1E} + N_{2E} \approx N_E, \quad N_{1Y} \approx N_Y, \tag{5}$$

The properties of the two magnetic plasmon resonance modes, such as the quality factor (obtained by measuring the rate of exponential decay of the electromagnetic energy for the given resonance mode), group velocity, effective mode volume, and normalized spatial intensity distributions, are calculated in our simulations and listed in Table 1. The other material parameters used in the calculation are also included in Table 1. In general, the $Yb^{3+}$ concentration is an order of magnitude higher than $Er^{3+}$ concentration. In our calculations, we fix the $Yb^{3+}$ concentration at $5.0 \times 10^{27} ions/m^3$ and predict the threshold $Er^{3+}$ concentrations. Substituting the pumping mode and lasing mode parameters into the equations (1-5), and assuming the total photon number of lasing mode in the cavity to be 1 and the pumping rate to be enough high, we can predict the $Er^{3+}$ threshold doping concentrations to be $4.6 \times 10^{25} ions/m^3$. It is worth to mention that, in our calculations, the metallic film is treated as bulk silver material with the $\tau$ equal to 33fs. However, the electron mean free path in metallic film is shorter than that in bulk material because of reflections at the film boundaries and interface roughness. So, the relaxation rate value in metal films is larger and can be obtained by means of experimental measurement. In our theoretical work, we consider the film influence on the property of metal material by reducing the quality factor of lasing mode, which corresponds to a larger metallic losses in metallic film. Figure 4(a) shows the threshold doping concentration plotted as a function of modified quality factor Q of lasing mode. It can be seen from the figure, as the quality factor decreases, the threshold doping concentration increases. When Q is decreased from 50 to 15, the corresponding threshold $Er^{3+}$ concentration is increased from $4.6 \times 10^{25} ions/m^3$ to $5.1 \times 10^{26} ions/m^3$. Typical $Er^{3+}$ concentration in Er:Yb:YCOB is in the range of $10^{25} \sim 10^{27}$ ion/$m^3$. So, Er:Yb:YCOB can theoretically satisfy the operating condition of the metallic magnetic resonance nanolaser. Choosing $Er^{3+}$ concentration at $6.0 \times 10^{26} ions/m^3$ and $8.0 \times 10^{26} ions/m^3$, we calculate the corresponding absorbed threshold pumping power ($\hbar\omega_p \sigma_Y v_p F_p N_p N_Y / \eta_p$, $\eta_p$ is the quantum efficiency and approximately equal to 1) respectively, as shown in Fig. 4(b). From Fig. 4(b), at a fixed $Er^{3+}$ concentration, the absorbed threshold pumping power increases with the decrease of Q factor. This means that more metallic loss must be compensated by higher pumping power.

As for the designed structure, varying the polarization direction of the 980nm incident light can control the appearance of the resonance for the pumping light. So, we can compare the two threshold pumping powers from the conventional (pumping light dones't resonate with the cavity) and resonant pumping design (pumping light resonates with the cavity) and obtain the enhancement factor. In two numerical experiments, the structure is pumped by the same 980 nm cw light with a focused Gaussian spot diameter of 1800nm but with different polarization direction controlling the appearance of the resonance. Our calculations show the enhancement factor is about 3000. The physical picture of enhancement comes from when pumping light resonates with the cavity, the incident pumping power can be more easy to be coupled into the cavity, then the coupled pumping light in cavity is more easy to be absorbed by $Yb^{3+}$ resulting from the Purcell factor and large spatial overlaps each other among the pump energy, lasing mode and gain medium.

In summary, a metallic double magnetic plasmon resonance nanolaser has been proposed. Due to that both the pumping mode and the lasing mode are resonant in the cavity, the pumping efficiency can be enhanced greatly. The FDTD method and a set of rate equations are introduced to model the operation of the laser and predict the lasing condition. Results prove that the proposed double resonance laser can work as a compact nanolaser with quite low threshold value.

# Table

Table 1. Parameters used in the calculations

| Parameters | Values | Notes |
|---|---|---|
| $Q_l$ | 50 | Calculated by FDTD |
| $Q_p$ | 60 | Calculated by FDTD |
| $V_l$ | $1.6 \times 10^{-3}$ | Calculated by FDTD |
| $V_p$ | $4.0 \times 10^{-2}$ | Calculated by FDTD |
| $v_l$ | $2.3 \times 10^8 \, m/s$ | Calculated by FDTD |
| $v_p$ | $2.1 \times 10^8 \, m/s$ | Calculated by FDTD |
| $\sigma_E$ | $5.0 \times 10^{-25} \, m^2$ | See [18] |
| $\sigma_Y$ | $8.0 \times 10^{-25} \, m^2$ | See [18] |
| $\tau_{2E}$ | $5.0 \times 10^{-3} \, s$ | See [18] |
| $\tau_{2Y}$ | $2.6 \times 10^{-3} \, s$ | See [17] |
| C | $1.3 \times 10^{-23} \, m^3/s$ | See [17] |
| $k_1, k_2$ | $5.0 \times 10^{-21} \, m^3/s$ | See [19,20] |

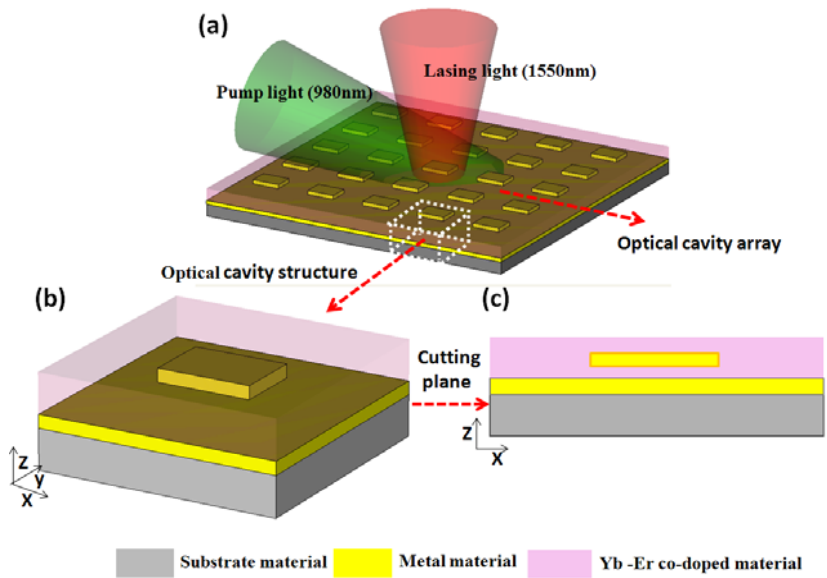

Figure 1. Metallic magnetic resonance nanolaser structures. (a) Laser arrays. (b) Single laser cell structure. (c) Cutting plane (z-x plane) of single laser cell structure.

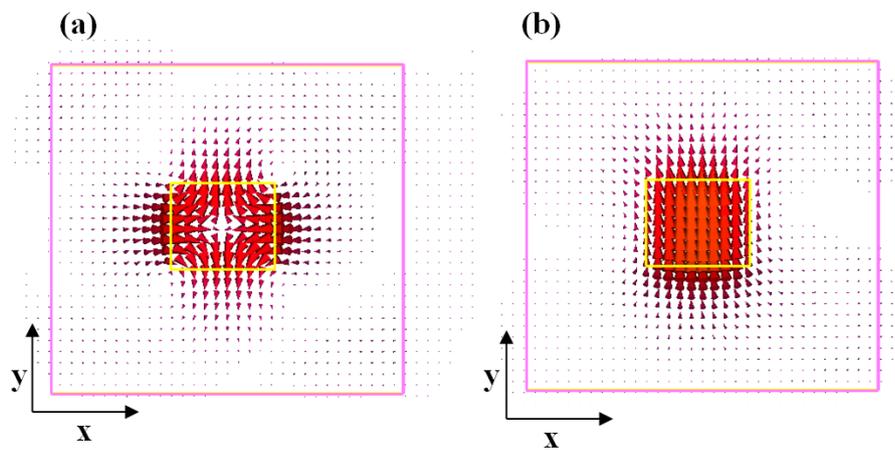

Figure 2. Magnetic field vectors of two magnetic resonance modes. (a) Corresponding to mode MP22. (b) Corresponding to mode $MP_{10}$.

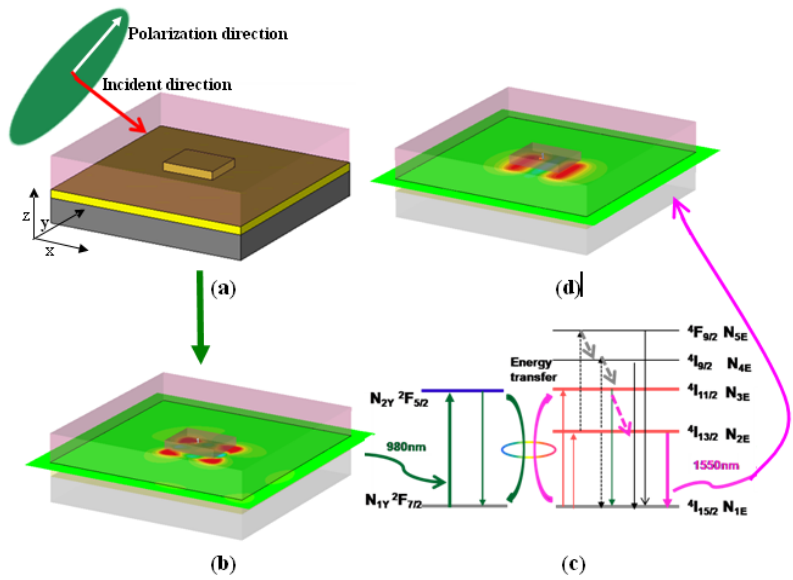

Figure 3. Schematic illustration of the operation principle of the laser. (a) Pumping light incident. (b) Resonantly pumping cavity mode(980nm). (c) Energy level diagram for the Er–Yb co-doped system. (d) Lasing cavity mode(1550nm).

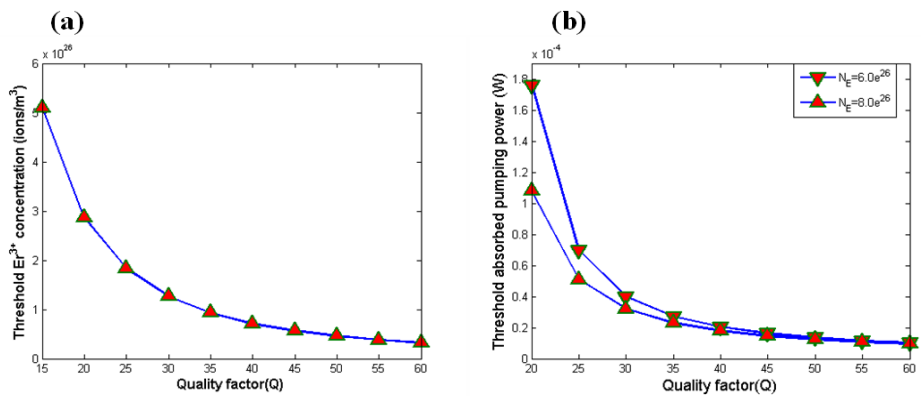

Figure 4. Threshold lasing condition. (a) Threshold $Er^{3+}$ concentration plotted as a function of modifiedquality factor Q. (b) Absorbed threshold pumping power plotted as a function of modified quality factor Q at two different erbium concentrations.